\documentclass[preprint,showpacs,superscriptaddress,prd]{revtex4}
\usepackage{graphicx,amsmath,dcolumn,bm}
\begin{document}
\title{\Large{\bf{Quark energy loss in semi-inclusive deep inelastic scattering of leptons on nuclei}}}
\author{Li-Hua Song }
\email[E-mail: ]{songlh@mail.heut.edu.cn}

\affiliation{Department of Physics, Hebei Normal
             University,
             Shijiazhuang 050016, P.R.China}
\affiliation{Hebei Advanced Thin Films Laboratory, Shijiazhuang
               050016, P.R.China}

\affiliation{ College of information,  Hebei Polytechnic
University, Tangshan 063009, P.R.China}

\author{Chun-Gui  Duan }
\email[E-mail: ]{duancg@mail.hebtu.edu.cn}

\affiliation{Department of Physics, Hebei Normal
             University,
             Shijiazhuang 050016, P.R.China}
\affiliation{Hebei Advanced Thin Films Laboratory, Shijiazhuang
               050016, P.R.China}



\begin{abstract}

Semi-inclusive deep inelastic scattering on nuclear targets is an
ideal tool to study the energy loss effect of an outgoing quark in
a nuclear medium. By means of the short hadron formation time, the
experimental data with the quarks hadronization occuring outside
the nucleus are picked out.  A leading-order analysis is performed
for the hadron multiplicity ratios as a function of the energy
fraction on helium, neon and copper nuclei relative to deuteron
for the various identified hadrons. It is shown that the nuclear
effects on parton distribution functions can be neglected. It is
found that the theoretical results considering the nuclear
modification of fragmentation functions due to quark energy loss
are in good agreement with the experimental data. Whether the
quark energy loss is linear or quadratic with the path length is
not determined. The obtained energy loss per unit length is
$0.38\pm0.03$ GeV/fm for an outgoing quark by the global fit.

\vskip 1.5cm

\noindent{\bf Keywords:} energy loss,quarks,hadron production,
formation time, fragmentation function.

\end{abstract}

\pacs{ 24.85.+p ; 
       25.30.-c;
       13.87.Fh; 
       12.38.-t;
       }

\maketitle
\newpage
\vskip 0.5cm

\section{Introduction}
A quantitative understanding of the quark propagation and hadron
formation processes in a nuclear medium would greatly benefit the
study of the quark-gluon plasma and its evolution in time. Quark
propagation in a nuclear medium involves competing processes like
hadronization of quarks and quark energy loss through multiple
scattering and gluon radiation. Semi-inclusive deep inelastic
scattering of leptons on the nucleus provides a unique opportunity
to study these effects on quark propagation and hadronization.
Therefore, in the past three decades, semi-inclusive deep
inelastic scattering on nuclear targets has been one of the most
active frontiers in nuclear physics and particle physics.

The pioneering measurements of the hadronization in a nuclear
medium were done at SLAC$^{[1]}$ in 1978.  Additional data on
multiplicity ratios were measured by the E665 experiment at
Fermilab$^{[2]}$ and the EMC experiment$^{[3]}$ using
ultra-high-energy muons on various nuclear targets. In 1991, the
EMC collaboration compared the differential multiplicities of
forward produced hadrons in deep inelastic muon scattering on
carbon, copper and tin targets with those from deuterium.
Multiplicity ratios were observed to increase toward unity as the
virtual photon energy $\nu$ and decrease with the energy fraction
$z$ of the virtual photon carried away by the leading hadron. More
recently, a series of semi-inclusive deep-inelastic scattering
measurements on various targets was performed  with the HERMES
detector at the DESY laboratory using a 27.6-GeV positron or
electron beam$^{[4-6]}$. The hadron multiplicity ratios on several
nuclei relative to the deuteron revealed a systematic decrease
with the mass number A and increase (decrease) with increasing
values of $\nu(z)$.

In semi-inclusive deep inelastic scattering of leptons on the
nucleus, there are two different nuclear effects,that is, the
initial state nuclear effect and final-state nuclear effect. In
the initial state, the lepton interacts with the quark in the
nucleus. The impact of the nuclear medium on parton distribution
functions is referred to as the initial state nuclear effect. In
the final state, the struck quark traverses the nuclear matter,
then fragments into hadrons. The so-called final-state nuclear
effect is the effect of the nuclear medium on the quark
propagation and hadron formation processes.

After the discovery of the EMC effect$^{[7]}$,  nuclear
modifications relative to nucleon parton distribution functions
are usually referred to as nuclear effects on parton distribution
functions, which include nuclear shadowing, antishadowing, the EMC
effect, and Fermi motion effect in different regions of the parton
momentum fraction. So far, three groups have presented their
global analysis of the nuclear parton distribution functions
analogous to those of the free proton: Eskola et al. $^{[8,9]}$,
by Hirai et al.$^{[10-12]}$, and by de Florian and Sassot (nDS)
$^{[13]}$.

Many theoretical phenomenological models were proposed to describe
the final-state nuclear effect in semi-inclusive deep inelastic
scattering of leptons on the nucleus. The phenomenological
models$^{[14-17]}$ used various formation times and absorption
cross sections for the various hadrons in the nuclear medium. The
gluon bremsstrahlung model$^{[18]}$ described the production of
fast leading mesons in nuclei by combination of a modification of
the fragmentation function in the medium owing to energy loss
caused by gluon radiation with hadron rescattering. The rescaling
model$^{[19,20]}$ hypothesized that the modification of quark
distribution and fragmentation functions was supplemented by
nuclear absorption. Another class of models$^{[21-25]}$ took
account of the energy loss that the struck quark experienced in
the nuclear environment without nuclear absorption effect. The
probabilistic coupled-channel transport model$^{[26,27]}$ was
based on the Boltzmann-Uehling-Uhlenbeck equation. These models
reproduce qualitatively the global features of the data. But it is
worth noting that the models do not perform quantitative fits to
experimental data by $\chi^2$ calculation.

Based on the Bialas picture$^{[28]}$, a more simplistic
presentation of the process of hadronization can be described in
terms of just the hadron formation time. The hadron formation time
is defined as the time between the moment that the quark is struck
by the virtual photon and the moment that the hadron is formed.
According to the hadron formation time, the quark hadronization
that occurs outside the nucleus can be distinguished from that
occuring inside the nucleus. If hadronization takes place outside
the nucleus, semi-inclusive deep inelastic lepton-nucleus
collisions are used to study the quark propagation in a nuclear
medium and, further, gain information on quark energy loss.

The nuclear Drell-Yan process is an ideal tool to investigate the
initial-state energy loss effect because the produced lepton pair
does not interact strongly with the partons in the nucleus. In the
previous articles$^{[29-32]}$, by using the nuclear parton
distribution functions from a global analysis, nuclear Drell-Yan
production cross-section ratios were calculated for 800-GeV
protons incident on a variety of nuclear targets, with
consideration of the energy loss of the projectile at the hadron
level and parton level separately. With the introduction of two
representative parametrizations of quark energy loss, we obtained
the incoming quark energy loss per unit length by comparing the
nuclear Drell-Yan experimental data. In this paper, the
experimental data with the quarks hadronization occuring outside
the nucleus are picked out in accordance with the hadron formation
time. The quark energy loss in semi-inclusive deep inelastic
lepton-nucleus scattering is extracted by the $\chi^2$ analysis
method. It is hoped to gain new knowledge about quark energy loss
in a nuclear medium.

The outline of the paper is as follows. A brief formalism for
 hadron production in semi-inclusive
deep inelastic scattering on the nucleus and nuclear modification
of the fragmentation functions owing to quark energy loss is given
in Sect.II, followed by a presentation of the formation time in
Sect.III. The method adopted in this analysis is described in
Sect.IV. The results and discussion obtained are presented in
sect.V. Finally, a summary is given in sect.VI.

\section{Hadron production in semi-inclusive deep inelastic
scattering on nuclei }

At leading order(LO) in perturbative QCD, the differential cross
section from hadron production in semi-inclusive deep inelastic
scattering of the lepton on the nucleus is written as
\begin{equation}
\frac{d\sigma^{3}}{dxdzd
\nu}=\sum_{f}e^{2}_{f}q^{A}_{f}(x,Q^{2})\frac{d\sigma^{lq}}{dxd\nu}D^{A}_{f|
h}(z,Q^{2}),
\end{equation}
where $e_f$ is the charge of the quark with flavor $f$,
$q^{A}_{f}(x,Q^{2})$ is the nuclear quark distribution function
with Bjorken variable $x$ and photon virtuality $Q^{2}$,
${d\sigma^{lq}}/{dxd\nu}$ is the differential cross section for
lepton-quark scattering at leading order, and $D^{A}_{f|
h}(z,Q^{2})$ is the nuclear modified fragmentation function of a
quark of flavour $f$ into a hadron $h$.

The hadron multiplicity is obtained from normalizing the
semi-inclusive deep inelastic scattering yield $N^{h}_{A}$ to the
deep inelastic scattering yield $N^{DIS}_{A}$,
\begin{equation}
\frac{1}{N^{DIS}_{A}}\frac{dN^{h}_{A}}{dzd
\nu}=\frac{1}{\sigma^{lA}}\int
dx\sum_{f}e^{2}_{f}q^{A}_{f}(x,Q^{2})\frac{d\sigma^{lq}}{dxd\nu}D^{A}_{f|
h}(z,Q^{2}),
\end{equation}
\begin{equation}
\sigma^{lA}=\int
dx\sum_{f}e^{2}_{f}q^{A}_{f}(x,Q^{2})\frac{d\sigma^{lq}}{dxd\nu},
\end{equation}
\begin{equation}
\frac{d\sigma^{lq}}{dxd\nu}=Mx\frac{4\pi\alpha_{s}^{2}}{Q^{4}}[1+(1-y)^{2}].
\end{equation}
where the integral range is determined according to the relative
experimental kinematic region, and $ \alpha _{s}$ and $y$ are the
fine structure constant and the fraction of the incident lepton
energy transferred to the target, respectively.

Furthermore, the hadron multiplicity ratio $R^{h}_{A/D}$ for the
identified hadron h on nucleus A relative to the deuteron is
defined as
\begin{equation}
R^{h}_{A/D}(\nu,z)=\frac{1}{N^{DIS}_{A}}\frac{dN^{h}_{A}(\nu,z)}{dzd\nu}
\Bigg/\frac{1}{N^{DIS}_{D}}\frac{dN^{h}_{D}(\nu,z)}{dzd\nu}.
\end{equation}

The struck quark can lose its energy owing to multiple scattering
and gluon radiation while propagating through the nucleus. The
quark energy fragmenting into a hadron shifts from $E_{q}=\nu$ to
$E'_{q}=\nu-\Delta E$, which results in a rescaling of the energy
fraction of the produced hadron:
\begin{equation}
z=\frac{E_{h}}{\nu}  \longrightarrow  z'=\frac{E_{h}}{\nu-\Delta
E},
\end{equation}
where $E_{h}$ and $\Delta E$ are, respectively, the measured
hadron energy and the quark energy loss in the nuclear medium. In
view of the rescaled energy fraction of the produced hadron, the
fragmentation function in the nuclear medium:
\begin{equation}
D^{A}_{f|h}(z,Q^{2})=D_{f|h}(z',Q^{2}),
\end{equation}
where $D_{f|h}$ is the standard (vacuum) fragmentation function of
a quark of flavour $f$ into a hadron $h$.

In our previous article$^{[32]}$, two parametrizations were
introduced for the quark energy loss. One is written as
\begin{equation}
\Delta E=\alpha <L>_{_{A}}.
\end{equation}
The other is presented as
\begin{equation}
\Delta E=\beta <L>_{A}^{2}.
\end{equation}
Here, $\alpha and \beta$ are the parameters that can be extracted
from experimental data. $<L>_{A}$ is the average path length
travelled by the quark in the nuclear matter. The two
parametrizations are called the linear and the quadratic quark
energy loss, respectively. If hadronization occurs well after the
hard quark has escaped from the nucleus, the length $<L>_{A}$ can
be given by
\begin{equation}
<L>_{A}=\frac{3}{4}R_{A},
\end{equation}
where the nuclear radius $R_{A}\simeq 1.12A^{1/3}$fm. If hadrons
are produced inside the medium, the length given previously is no
longer correct. Therefore, excluding the influence of nuclear
absorption, and considering only the case where hadrons are
produced outside the nucleus, the hadron multiplicity can be
expressed as
\begin{equation}
\frac{1}{N^{DIS}_{A}}\frac{dN^{h}_{A}}{dzd
\nu}=\frac{1}{\sigma^{lA}}\int
dx\sum_{f}e^{2}_{f}q^{A}_{f}(x,Q^{2})\frac{d\sigma^{lq}}{dxd\nu}D_{f|h}(z',Q^{2}).
\end{equation}

\section{ The hadron formation time}
To pick out the experimental data with quark hadronization
occurring outside the nucleus, the characteristic physics variable
is the hadron formation time $t$. According to the hadron
formation time, it can be determined whether a hadron is produced
outside or inside the nucleus. If $t
>\frac{3}{4}R_{A}$, hadronization occurs outside the nucleus.
Otherwise, hadrons are produced inside the nucleus.

There are many expressions of hadron formation time. Three
representative parametrizations are as following. The first is
based on the Lund model and extracted from HERMES data$^{[33]}$,
\begin{equation}
t_{H}=z^{0.35}(1-z)\nu/\kappa.
\end{equation}
where $\kappa$ is the string tension (string constant) with
numerical value $\kappa = 1GeV/fm$. The second is given by Bialas
and Gyulassy$^{[34]}$ following from the Lund string model,
\begin{equation}
t_{Lund}=[\frac{ln(1/z^{2})-1+z^{2}}{1-z^{2}}]\times\frac{z\nu}{k}.
\end{equation}
The third expression is composed of the characteristic formation
time of the hadron in its rest frame $\tau_{0}$ and the Lorentz
factor$^{[28]}$,
\begin{equation}
t_{Lor}=\tau_{0}\frac{z\nu}{m_{h}},
\end{equation}
where $m_{h}$ is the mass of the hadron $h$. In Ref.[26],
$\tau_{0}=0.5fm$ and $m_{h}=0.14GeV$ for the pion meson.

\begin{figure}[!]
\centering
\includegraphics*[width=90.3mm,height=9cm]{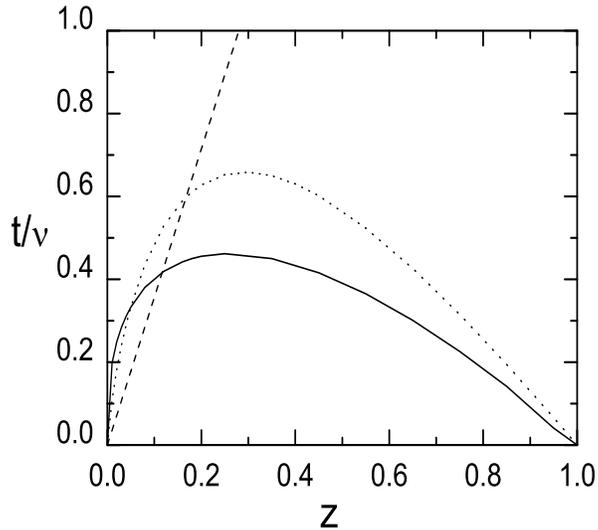}
\vspace{-0.25cm} \caption{$t/\nu$ as a function of $z$ for the
three different parametrizations of the hadron formation time.
Solid, dotted, and dashed curve indicate hadron formation times
$t_{H}$, $t_{Lund}$, and $t_{Lor}$, respectively.}
\end{figure}

In Fig.1, the ratios $t/\nu$ as a function of $z$ for the three
formation time expressions are presented  by the solid curve
($t_{H}$), dotted curve ($t_{Lund}$), and dashed curve
($t_{Lor}$), respectively. As shown in Fig.1, the formation time
$t_{H}$ is the shortest at $z > 0.12$ when $\nu$ is kept constant.

\section{ The analysis method for experimental data}

To extract the value of the parameter in the quark energy loss
expression, ignoring the correlations in the measurement errors,
the chi-square function for each experimental sample $l$ with $m$
data points is defined as$^{[35]}$
\begin{equation}
\chi^{2}_{l}=\sum_{i}^{m} \Bigg[\frac{R^{h, data}_{A/D, i}-R^{h,
theo}_{A/D, i}}{R^{h, err}_{A/D, i}} \Bigg]^2,
\end{equation}
where $R^{h, data}_{A/D, i}$ and $R^{h, theo}_{A/D, i}$ indicate
separately the experimental data and theoretical values of the
hadron multiplicity ratio $R^{h}_{A/D}$. $R^{h, err}_{A/D, i}$
represents the statistical and the uncorrelated systematic errors,
added in quadrature.

The calculated values of the parameters in quark energy loss
expressions are determined from the minimization of the
$\chi^{2}_{l}$ function,
\begin{equation}
\chi^{2}_{l, min}(\alpha_l,\beta_l)\equiv
min[\chi^{2}_{l}(\alpha_l,\beta_l)],
\end{equation}
with the experimental data sample $l$. One standard deviation of
the relevant parameter on the fitted value corresponds to an
increase of $\chi^{2}_{l}$ by 1 unit from its minimum
$\chi^{2}_{l,min}$:
\begin{equation}
\Delta
\chi^{2}_{l}=\chi^{2}_{l}(\alpha_{l}\pm\delta\alpha_{l})-\chi^{2}_{l,min}=1,
\end{equation}
\begin{equation}
\Delta
\chi^{2}_{l}=\chi^{2}_{l}(\beta_{l}\pm\delta\beta_{l})-\chi^{2}_{l,min}=1,
\end{equation}
where $\delta\alpha_{l}$ and $ \delta\beta_{l}$ are the
uncertainties of the fitted parameters.

According to the treatment method of the Particle Data
Group$^{[36]}$, if $\chi^{2}_{l, min}/(m-1)$ is $\leq1$, the
results on the theoretical values of the parameters and their
uncertainties can be accepted.  When $\chi^{2}_{l, min}/(m-1)$ is
$>1$, but not greatly so, a scale factor $S_{l}$ is defined as
\begin{equation}
 S_{l}=\sqrt{\chi^{2}_{l, min}/(m-1)}.
\end{equation}
The uncertainty of the fitted parameter from 1 standard deviation
is rescaled,
\begin{equation}
\delta\alpha^S_{l}=S_{l}\times\delta\alpha_{l},\hspace{1cm}
\delta\beta^S_{l}=S_{l}\times\delta\beta_{l}.
\end{equation}
If $\chi^{2}_{l, min}/(m-1)\leq 1$, the scale factor $S_{l}=1$.

When there are $N$ experimental data samples, the parameters in
quark energy loss expressions can be determined from a global fit
of all used data by means of minimization of the $\chi^{2}$
function,
\begin{equation}
\chi^{2}_{min}(\alpha,\beta)=\sum_{l=1}^{N}S_{l}^{-1}\chi_{l}^{2}(\alpha,\beta).
\end{equation}
The uncertainty of the fitted parameter can be calculated on the
analogy of the single experimental sample.

\section{ Results and discussion }

To investigate the quark energy loss in semi-inclusive deep
inelastic scattering of leptons on the nucleus, and determine the
values of the parameters $\alpha$ and $\beta$ in quark energy loss
expressions, experimental data with quark hadronization occuring
outside the nucleus are picked out by means of the short hadron
formation time $t_{H}$.  For the hadron multiplicity ratio
$R^{h}_{A/D}$ as a function of the energy fraction of the virtual
photon carried away by the leading hadron,  the kinematical ranges
of selected experimental data are $0.16 \leq z \leq0.75$ for a
helium target, $0.16 \leq z \leq 0.45$ for a neon target from the
HERMES experiment$^{[6]}$ with $6.0 GeV<  \nu < 23.5GeV$, and
$0.07 \leq z \leq 0.54$ and $10 GeV<  \nu < 230GeV$ for a copper
target from the EMC experiment$^{[3]}$, respectively.

Neglecting the nuclear effects in semi-inclusive deep inelastic
scattering of leptons on the nucleus, the hadron multiplicity
ratios $R^{h}_{A/D}(z)$ are calculated by combining the CTEQ6L
parton density in the proton$^{[37]}$ with the vacuum
fragmentation functions from the leading-order analysis of
$e^+e^-$ data performed by Kretzer$^{[38]}$. It is found that the
hadron multiplicity ratios are equal to 1. Considering only the
initial nuclear effect, by means of the nuclear distribution
functions$^{[12]}$ and vacuum fragmentation functions, the
calculated $R^{h}_{A/D}(z)$ values are approximately 1. It is
demonstrated that there are not nuclear effects on the parton
distribution functions in semi-inclusive deep inelastic
scattering. This result is in accordance with the theoretical
prediction. In HERMES and EMC experiments, the  minimal value of
the Bjorken variable $x$ is approximately $0.02$. Because the
(nuclear) parton distribution functions appear both in the
numerator and in the denominator of the hadron multiplicity ratio,
mathematical integration for the Bjorken variable results in the
nuclear effects on the parton distribution functions being
negligible in the region $x>0.02$. Therefore, we pay attention
only to the final-state nuclear effect in semi-inclusive deep
inelastic scattering.

As for the final state nuclear effect in semi-inclusive deep
inelastic scattering of leptons on the nucleus, the struck quark
traverses the nuclear matter, loses its energy owing to multiple
scattering and gluon radiation, then fragments into a hadron. If
the produced hadron is formed outside the nucleus, semi-inclusive
deep inelastic scattering  on nuclear targets provides a good
opportunity to investigate the energy loss effect of fast quarks
propagating in the nuclear medium. By using the CTEQ6L parton
density in the proton$^{[37]}$  together with the vacuum
fragmentation functions $^{[38]}$, meanwhile taking account of the
quark energy loss in the final-state effect, the hadron
multiplicity ratios $R^{h}_{A/D}(z)$ at leading order are
calculated and compared with the experimental data with quark
hadronization occuring outside the nucleus. In our calculation,
the integral regions of the variable $\nu$ are $6.0 GeV<  \nu <
23.5GeV$ for the HERMES data and $10 GeV<  \nu < 230GeV$ for EMC
data.
\begin{table}[t,m,b]
\caption{The values of parameters $\alpha$ and $\beta$ and
$\chi^{2}/ndf$ from a fit to the selected data on
$R^{h}_{A/D}(z)$.}
\begin{ruledtabular}
\begin{tabular*}{\hsize}
{c@{\extracolsep{0ptplus1fil}} c@{\extracolsep{0ptplus1fil}}
c@{\extracolsep{0ptplus1fil}}}
         Hadron  & $\alpha(\chi^{2}/ndf)$ &$\beta(\chi^{2}/ndf) $\\
         \colrule
       $\pi^{+}$   & $0.35\pm0.08$(0.81) & $0.14\pm0.06(0.80)$\\
       $\pi^{-}$ &$0.35\pm0.06$(0.62)&$0.14\pm0.06(0.61)$\\
       $k^{+}$   & $0.46\pm0.12$(0.49)&$0.21\pm0.07(0.48)$\\
     $k^{-}$ &$0.78\pm0.16$(0.41)&$0.36\pm0.08(0.39)$\\
\end{tabular*}
\end{ruledtabular}
\end{table}

\begin{table}[t,m,b]
\caption{The values of parameters $\alpha$ and $\beta$ and
$\chi^{2}/ndf$ from a fit to the selected data on $R^{h}_{A/D}(z)$.}
\begin{ruledtabular}
\begin{tabular*}{\hsize}
{c@{\extracolsep{0ptplus1fil}} c@{\extracolsep{0ptplus1fil}}
c@{\extracolsep{0ptplus1fil}}}
         Hadron  & $\alpha(\chi^{2}/ndf)$ &$\beta(\chi^{2}/ndf) $\\
         \colrule
        $\pi^{+}$ & $0.43\pm0.06$(0.14)&$0.135\pm0.020(0.13)$\\
      $\pi^{-}$&$0.39\pm0.06$(0.42)&$0.12\pm0.02(0.42)$\\
     $k^{+}$ & $0.31\pm0.08$(0.05) &$0.10\pm0.03(0.05)$\\
       $k^{-}$&$0.51\pm0.06$(0.52)&$0.165\pm0.020(0.48)$\\
\end{tabular*}
\end{ruledtabular}
\end{table}

\begin{table}[t,m,b]
\caption{The values of parameters $\alpha$ and $\beta$ and
$\chi^{2}/ndf$ from a fit to the selected data on $R^{h}_{A/D}(z)$.}
\begin{ruledtabular}
\begin{tabular*}{\hsize}
{c@{\extracolsep{0ptplus1fil}} c@{\extracolsep{0ptplus1fil}}
c@{\extracolsep{0ptplus1fil}}}
         Hadron  & $\alpha(\chi^{2}/ndf)$ &$\beta(\chi^{2}/ndf) $\\
         \colrule
  h &$0.24\pm0.03$(1.63) &$0.054\pm0.009(1.64)$\\
\end{tabular*}
\end{ruledtabular}
\end{table}

\begin{table}[t,m,b]
\caption{The values of parameters $\alpha$ and $\beta$ and
$\chi^{2}/ndf$ from a fit to the selected data on $R^{h}_{A/D}(z)$.}
\begin{ruledtabular}
\begin{tabular*}{\hsize}
{c@{\extracolsep{0ptplus1fil}} c@{\extracolsep{0ptplus1fil}}
c@{\extracolsep{0ptplus1fil}}}
         Exp. &  $\alpha(\chi^{2}/ndf)$ &$\beta(\chi^{2}/ndf) $\\
         \colrule
     HERMES(He)
       & $0.41\pm0.07(0.78)$&$0.19\pm0.03(0.78)$\\
           HERMES(Ne)
      &$0.42\pm0.03$(0.52)&$0.133\pm0.010$(0.52)\\

      EMC(Cu) &$0.24\pm0.03$(1.63) &$0.054\pm0.009(1.64)$\\

  Global analysis &$0.38\pm0.03$(0.86)  &$0.125\pm0.002(1.07)$\\

\end{tabular*}
\end{ruledtabular}
\end{table}

Table I.,Table II. and Table III. summarize $\chi^2$ per number of
degrees of freedom ($\chi^2/ndf$) and the determined parameters
$\alpha$ and $\beta$ in quark energy loss expressions by
calculating the hadron multiplicity ratios $R^{h}_{A/D}(z)$ on
helium, neon and copper nuclei relative to the deuteron for
various identified hadrons in the HERMES$^{[6]}$ and EMC
experiments$^{[3]}$. It is shown that the theoretical results
considering the nuclear modification of fragmentation functions
owing to quark energy loss are in good agreement with the
experimental data. The calculated results from the linear and
quadratic quark energy loss are in favor of the attenuation of
hadron multiplicity ratios $R^{h}_{A/D}(z)$. Therefore, the hadron
multiplicity ratios $R^{h}_{A/D}(z)$ can not determine whether the
quark energy loss is linear or quadratic with the path length. The
nuclear Drell-Yan differential cross-section ratios for heavy
nuclei versus deuterons at a lower incident proton energy are
hoped to distinguish between the linear and the quadratic
dependence of quark energy loss$^{[30]}$.

\begin{figure}[t,m,b]
\centering
\includegraphics*[width=140.3mm,height=11cm]{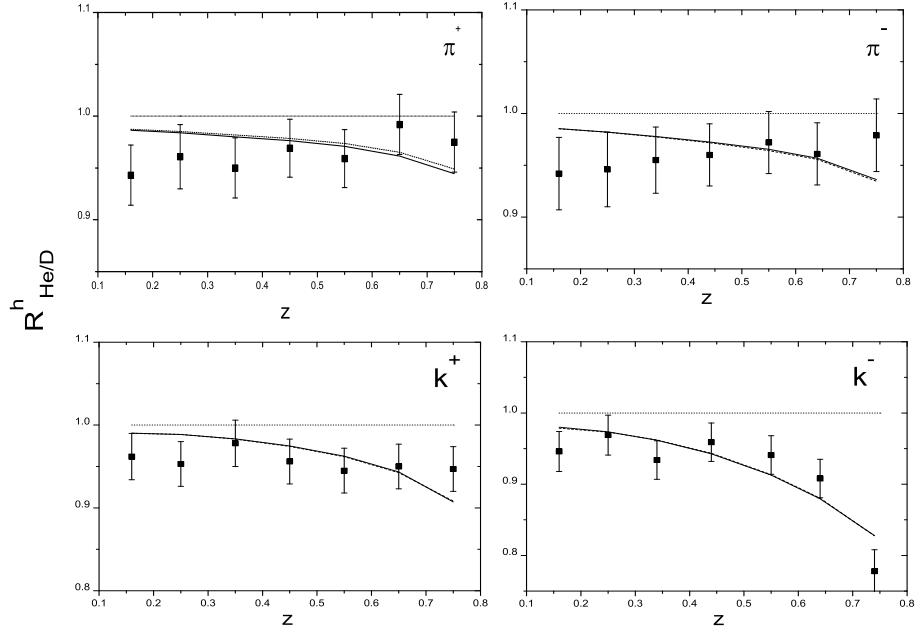}
\vspace{-0.25cm} \caption{The multiplicity ratios $R^{h}_{He/D}$
for identified hadron $k^{-}$, $k^{+}$, $\pi^{-}$ and $\pi^{+}$ as
a function of $z$. The solid and dashed curves correspond to the
results from the linear and quadratic quark energy loss,
respectively. The experimental data are taken from HERMES
data$^{[6]}$. The error bars represent the systematic
uncertainty.}
\end{figure}

\begin{figure}[t,m,b]
\centering
\includegraphics*[width=140.3mm,height=11cm]{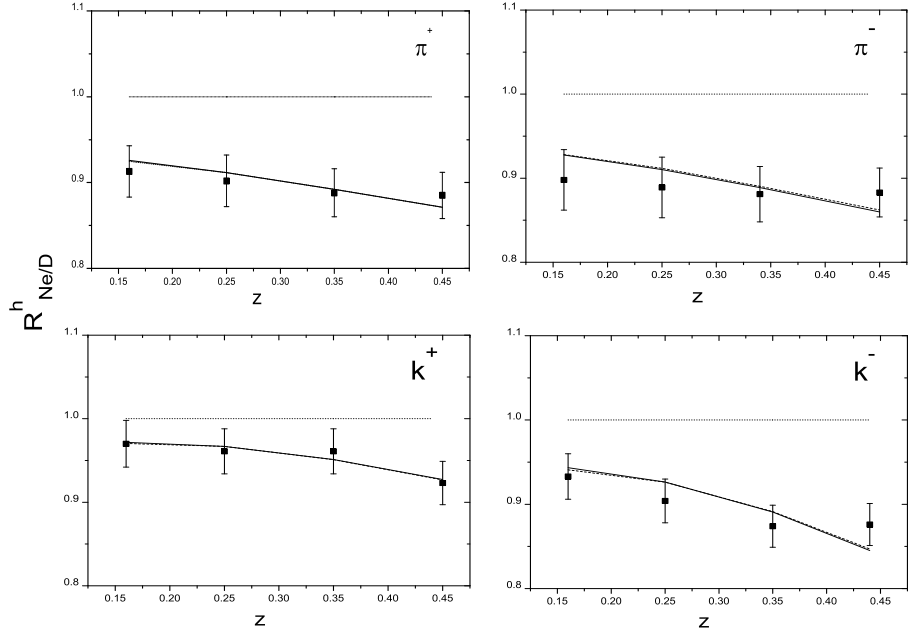}
\vspace{-0.25cm} \caption{The multiplicity ratios $R^{h}_{Ne/D}$
for identified hadron $k^{-}$, $k^{+}$, $\pi^{-}$ and $\pi^{+}$.
The comments are the same as Fig.2.}
\end{figure}

\begin{figure}[!]
\centering
\includegraphics*[width=90.3mm, height=8cm]{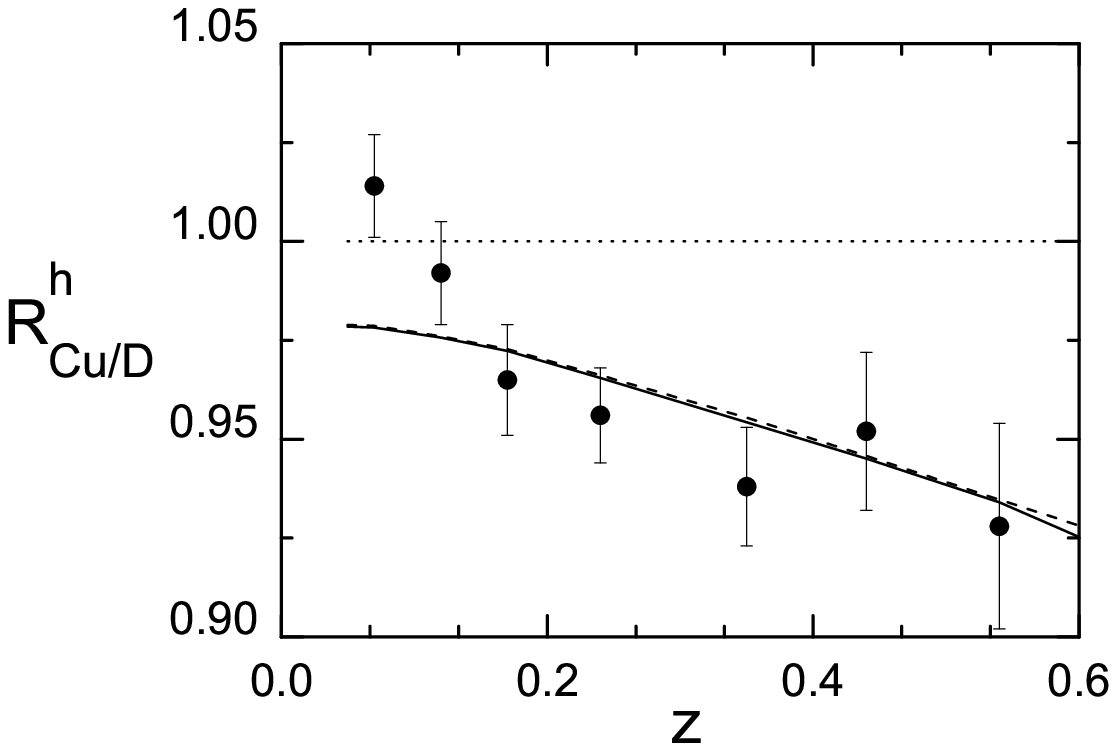}
\vspace{-0.5cm} \caption{The multiplicity ratios $R^{h}_{Cu/D}$
for hadron $h$ as a function of $z$. The experimental data are
selected from EMC data$^{[3]}$. The comments are the same as
Fig.2.}
\end{figure}

Fig.2, Fig.3 and Fig.4 show, respectively, the hadron multiplicity
ratios $R^{h}_{A/D}(z)$ on three nucleus versus deuteron for the
various identified hadrons. The solid and dashed curves are the
results on hadron multiplicity ratios $R^{h}_{A/D}(z)$ from the
linear and  quadratic quark energy loss, respectively. As can be
seen, the solid curves have almost no difference with the dashed
curves. In nuclear medium, the rescaled energy fraction is bigger
than the energy fraction in vacuum fragmentation function. Because
the vacuum fragmentation function decreases with the increase of
energy fraction $z$, the hadron multiplicity ratio
$R^{h}_{A/D}(z)$ decreases with the energy fraction as presented
in figures. As for HERMES experimental data, it is found that as
the target nucleus A becomes heavier, the nuclear suppression of
hadron multiplicity ratio $R^{h}_{A/D}(z)$ becomes bigger. The
result originates from the quark energy loss effect. In the quark
energy loss expressions given above, the magnitude of quark energy
loss in nuclear medium has close relation with the atomic number
A. The quark  energy loss becomes bigger with the increase of the
atomic number A. With taking example for the identified hadron
$k^-$, the suppression of $R^{h}_{A/D}(z)$ due to quark energy
loss effect is approximately $2\%$ to $7\%$ for helium target  and
$6\%$ to $20\%$ for neon target  in the range $0.1\leq z \leq0.5$,
respectively.

By following the $\chi^2$ analysis method described in Sect.IV,
the values of $\alpha$ and $\beta$ in quark energy loss
expressions can also be extracted from a global fit of all
available experimental data with the quarks hadronization occuring
outside the nucleus. The values of $\alpha$ and $\beta$ extracted
from the individual fits of each data sample, as well as their
corresponding (rescaled) error, $\chi^2/ndf$ are summarized in
Table IV.

As for the linear quark energy loss, the mean energy loss per unit
length  is $dE/dL=0.38\pm 0.03GeV/fm$, which is smaller than
$dE/dL=0.5 GeV/fm $ extracted by E.Wang and X-N. Wang$^{[23]}$
through comparing qualitatively  with  the hadron multiplicity
ratio for neon and krypton target from HERMES  data. The
overestimation from Ref.[23] of quark energy loss is due to
neglecting the nuclear absorption in semi-inclusive deep inelastic
scattering of lepton on nucleus.

On the basis of theoretical research$^{[39]}$, the mean energy
loss of an outgoing parton is 3 times larger than that in the case
of partons approaching the medium,
$({dE}/{dL})_{out}=3({dE}/{dL})_{in}$. Because of using the
experimental data from semi-inclusive deep inelastic scattering of
lepton on nuclei, the obtained here $dE/dL=0.38\pm 0.03GeV/fm$ is
the mean energy loss of an outgoing quark. In our previous
article$^{[32]}$, the energy loss of the projectile quark was
extracted from the nuclear Drell-Yan production  for 800GeV
protons incident on a variety of nuclear targets. It was shown
that the mean energy loss of an incoming quark is $dE/dL=1.26
GeV/fm$. It is worth emphasizing that the obtained energy loss of
an incoming quark depends strongly on the nuclear parton
distribution functions(see Ref.[32] for more detail discussion).
It is apparent that our result does not agree with the theoretical
prediction on the relation between the energy loss of incoming
quark and that of outgoing quark. Therefore, it is desirable to
operate precise measurements on the nuclear Drell-Yan reactions at
lower incident proton energy and semi-inclusive deep inelastic
scattering on nucleus at ultra-high lepton energy.

\section{ Summary }

The semi-inclusive deep inelastic scattering  on nuclear targets
is an ideal tool to  study energy loss effect of outgoing quark in
nuclear medium. A leading order analysis has been performed on the
hadron multiplicity ratio from the experimental data with the
quarks hadronization occuring outside the nucleus. It is shown
that there is not the nuclear effects on the parton distribution
functions in the semi-inclusive deep inelastic scattering. Our
results show that the theoretical results with the fragmentation
functions modified due to quark energy loss are in good agreement
with the experimental data. Whether the quark energy loss is
linear or quadratic with the path length does not be determined.
We obtain the energy loss per unit length $dE/dL=0.38\pm
0.03GeV/fm$ for an outgoing quark by the global fit to all
selected data. By combining our previous discussion on the nuclear
Drell-Yan process, the obtained energy loss of an incoming quark
and outgoing quark is not in support of the theoretical prediction
$({dE}/{dL})_{out}=3({dE}/{dL})_{in}$. Therefore, we desire to
perform precise measurements at J-PARC$^{[40]}$, Fermilab
E906$^{[41]}$ in the future. These new experimental data can
provide insight on the energy loss of an incoming quark
propagating in cold nucleus.

\vskip 1cm
{\bf Acknowledgments}
This work was supported in part by the National Natural Science
Foundation of China(10575028) and  Natural Science Foundation of
Hebei Province(A2008000137).

\end{document}